\documentclass[aps,pra,twocolumn,showpacs,superscriptaddress,10pt,longbibliography]{revtex4-2}

\usepackage[colorlinks = true,linkcolor = red,citecolor = magenta]{hyperref}
\usepackage[sort&compress]{natbib}
\usepackage{scalerel}
\usepackage[normalem]{ulem}
\usepackage{xcolor}
\usepackage{bm}
\usepackage{amsmath}
\usepackage{amsfonts}
\usepackage{amssymb}
\usepackage{graphicx}
\usepackage{subfigure}
\usepackage{mathtools}
\usepackage{physics}
\usepackage{dsfont}
\usepackage{float}
\usepackage{lmodern}
\usepackage[many]{tcolorbox}
\usepackage{empheq}
\usepackage{cleveref}
\usepackage{textcomp}

\begin{document}

\title{Conductance transition with interacting bosons in an Aharonov-Bohm cage}
\author{A. R. Kolovsky}
\affiliation{Kirensky Institute of Physics, Federal Research Center KSC SB RAS, 660036, Krasnoyarsk, Russia} 
\affiliation{School of Engineering Physics and Radio Electronics, Siberian Federal University, 660041, Krasnoyarsk, Russia} 
\affiliation{Center for Theoretical Physics of Complex Systems, Institute for Basic Science, 34126 Daejeon, Republic of Korea}
\author{P. S. Muraev}
\affiliation{Kirensky Institute of Physics, Federal Research Center KSC SB RAS, 660036, Krasnoyarsk, Russia}
\affiliation{School of Engineering Physics and Radio Electronics, Siberian Federal University, 660041, Krasnoyarsk, Russia}
\affiliation{IRC SQC, Siberian Federal University, 660041, Krasnoyarsk, Russia}
\author{S. Flach}
\affiliation{Center for Theoretical Physics of Complex Systems, Institute for Basic Science, 34126 Daejeon, Republic of Korea}
\date{\today}
\begin{abstract}
We study transport of interacting bosons through an Aharonov-Bohm cage - a building block of flat band networks - with coherent pump and sink leads.
In the absence of interactions the cage is insulating due to destructive interference.
We find that the cage stays insulating up to a critical value of the pump strength in the presence of mean field interactions, while the quantum regime  induces particle pair transport and weak conductance below the critical pump strength.
A swift crossover from quantum into the classical regime upon further pump strength increase is observed.
We solve the time dependent master equations for the density matrix of the many body problem both in the classical, pure quantum, and pseudoclassical regimes.
We start with an empty cage and switch on driving. 
We characterize the transient dynamics, and the complexity of the resulting steady states and attractors.
Our results can be readily realized using experimental platforms involving interacting ultracold atoms and photons on finetuned optical lattices.
\end{abstract}
\maketitle

{\em Introduction.}
Flatbands arise in the band structure of finetuned tigh-binding networks and are used in various setups in condensed matter physics and photonics \cite{bergholtz2013topological,derzhko2015strongly,Leykam:2018ab,Leykam:2018aa}. Particular interest has been paid to the case when all bands are flat - ABF lattices, as originally observed for particular magnetic flux values threading the lattice, and therefore coined Aharonov-Bohm cages (AB) due to the complete destructive interference induced trapping of noninteracting particles \cite{vidal1998aharonov,vidal2000interaction}. ABF lattices can be in general diagonalized with a finite number of local unitary transformations, support compact localized eigenstates (CLS), and serve as the starting ground for a variety of single particle and many-body perturbations which lead to different nonperturbative phases of matter \cite{tovmasyan2013geometry,tovmasyan2018preformed,kuno2020flat,kuno2020interaction,danieli2020many,danieli2021nonlinear,danieli2021quantum,vakulchyk2021heat,orito2021nonthermalized,cadez2021metal,Kim:2022uw,Lee2023Critical}.

\begin{figure}
\includegraphics[width=8.5cm,clip]{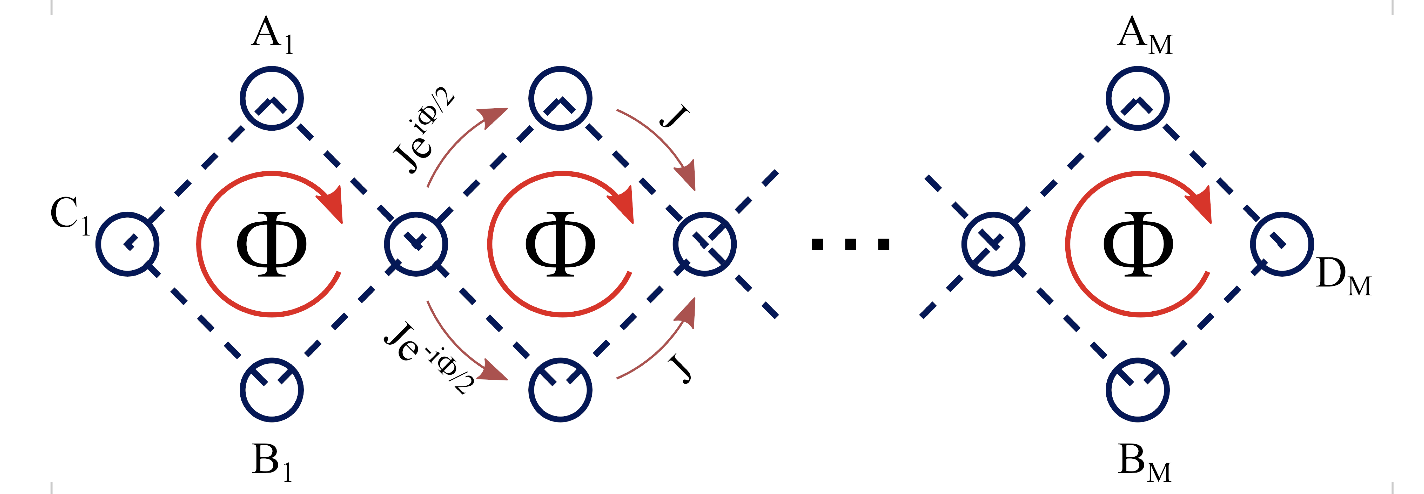}
\caption{The flux rhombic lattice (diamond chain). The all bands flat case corresponds to $\Phi=\pi$.}
\label{fig1}
\end{figure}

A famous example of such an ABF lattice is the  $\pi$-flux rhombic lattice \cite{vidal2000interaction} (also known as the diamond lattice) which is shown in Fig.~\ref{fig1}. Here the term "$\pi$-flux" means that the sum of phases of the nearest-neighbour hopping matrix elements equals to $\pi$. All eigenstates of the quantum particle in this lattice are compact localized states which prohibit any transport across the lattice. This statement, however, is valid only for non-interacting particles. Finite inter-particle interaction will recover transport in general \cite{vidal2000interaction,danieli2021quantum,121}, though additional single particle and interaction finetuning can prohibit transport even in the interacting case \cite{danieli2021nonlinear,danieli2021quantum,danieli2020many,doi:10.1063/1.5041434}.

In the present work we address the effect of inter-particle interaction on the transport of Bose particles across the diamond lattice from the viewpoint of laboratory  experiments  where one injects  bosons into the first site of the lattice by using an external coherent driving pump and withdraws them from the last site with a sink. Nowadays such experiments can be performed by using different physical platforms, for example, super-conducting circuits \cite{Fitz17,Fedo21,Martinez2023}.  In short, one arranges interacting transmons (micro-resonators coupled to Josephson junctions) in a lattice,  drives the first transmon with a microwave field, and reads the signal off from the last transmon. The crucial  impact of Josephson junctions is that they introduce an effective inter-particle interaction for photons in the micro-resonators.  Hence, that setup can be modeled by the Bose-Hubbard Hamiltonian. Another promising system are photonic crystals \cite{Szam05}. In a recent laboratory experiment \cite{Cace22} the authors realized the $\pi$-flux rhombic lattice and proved the absence of transport for non-interacting photons. 

{\em Model.}
The above considered laboratory setup is described by the following master equation for the reduced density matrix $\widehat{{\cal R}}(t)$ of microwave photons,
\begin{equation}
\label{b1}
\frac{\partial \widehat{{\cal R}}}{\partial t}=-\frac{i}{\hbar}[\widehat{{\cal H}}, \widehat{{\cal R}}] -\frac{\gamma}{2}
\left(\hat{a}_L^{\dagger}\hat{a}_L\widehat{\cal R }-2\hat{a}_L\widehat{\cal R }\hat{a}_L^{\dagger}
+\widehat{\cal R }\hat{a}_L^{\dagger}\hat{a}_L \right) \;,
\end{equation}
where $\gamma$ is the rate of photon absorption by a measurement device and the Hamiltonian  $\widehat{{\cal H}}$ has the form
\begin{eqnarray}
\label{b2}
\widehat{{\cal H}}= \hbar \Delta \sum_{\ell=1}^{L}\hat{n}_{\ell}
-\sum_{\ell,\ell'=1}^L \frac{\hbar J_{\ell,\ell'}}{2}\left( \hat{a}_{\ell}^{\dagger}\hat{a}_{\ell'} +{\rm h.c.} \right) \\
\nonumber
+\frac{\hbar^2 g}{2}\sum_{\ell=1}^{L}\hat{n}_{\ell} (\hat{n}_{\ell}-1) 
+ \frac{\sqrt{\hbar}\Omega}{2}(\hat{a}_1^\dagger + \hat{a}_1) \;.
\end{eqnarray}
In Eqs.~(\ref{b1})-(\ref{b2}) $\hat{a}_\ell$ and $\hat{a}_\ell^\dagger$ are the standard annihilation and creation operators which commute to unity, $\hat{n}_\ell=\hat{a}_\ell^\dagger\hat{a}_\ell$ is the particle number operator, $\Omega$ is the Rabi frequency (which is proportional to the amplitude of the driving field), $\Delta$ the detuning of the driving frequency from the linear frequency of quantum oscillators, $g$ the macroscopic interaction constant (which determines the nonlinearity  $U=\hbar g$ of the energy spectrum of quantum oscillators), $J_{\ell,\ell'}$  are the hopping matrix elements (i.e., the couplings between oscillators), $L$ is the total number of sites, and $\hbar$ is the dimensionless Planck constant \footnote{Notice, that from the viewpoint of a laboratory experiment the classical limit $\hbar\rightarrow0$ corresponds to scaling the nonlinearity of the oscillator energy spectrum as $U=\hbar g$, and the driving strength as $\Omega \rightarrow \Omega/\sqrt{\hbar}$.}.
We label the rhomb sites as shown in Fig.~\ref{fig1} where $C_2\equiv D_1$, $C_3\equiv D_2$, etc., and we shall use the gauge where $J_{AD}=J_{BD}=J$ and $J_{CA}=-J_{CB}=iJ$.  In what follows we set $J=1$ which implies that all system parameters are measured in units of the hopping frequency $J$, and time in units of its inverse $1/J$.
The case of non-interacting particles results in destructive interference on the first $D$-site of the diamond lattice, which blocks all transport. Thus, to understand the effect of finite inter-particle interactions it suffices to consider the lattice consisting of a single rhomb.

{\em Mean field approach.}
First we analyze the classical (mean-field) problem by using the Gross-Pitaevskii equations on the rhomb lattice, i.e. by replacing annihilation and creation operators by c-numbers in the presence of damping and driving\cite{KordasWitthautBuonsanteVezzaniBurioniKaranikasWimberger2015}.   For the lattice consisting of a single rhomb we have 
\begin{eqnarray}
\label{1}
i\dot{C}=\left(\Delta+g|C|^2\right) C -\frac{i}{2}A+\frac{i}{2}B + \frac{\Omega}{2} \\
\nonumber
i\dot{A}=\left(\Delta+g|A|^2 \right)A +\frac{i}{2}C-\frac{1}{2}D \\
\nonumber
i\dot{B}=\left(\Delta+g|B|^2\right) B -\frac{i}{2}C-\frac{1}{2}D \\
\nonumber
i\dot{D}=\left(\Delta+g|D|^2\right) D -\frac{1}{2}A-\frac{1}{2}B - i\frac{\gamma}{2} D \;,
\end{eqnarray}
where $C,A,B,D$ are now time-dependent complex amplitudes of the local oscillators at the rhomb nodes and $\dot{X} \equiv dX/dt$. We use the Rabi frequency $\Omega$ as our control parameter and fix  $g=0.5$, $\gamma=0.2$ and $\Delta=-0.5$.
The above equations are invariant under a reflection symmetry $A \rightarrow -B$, $B \rightarrow -A$, with $D=0$. To enforce $D=0$ we need $A(t)=-B(t)$.
Such states are the generalization of a flatband CLS which persist due to the destructive interference of waves from $A$ and $B$ reaching site $D$ \cite{doi:10.1063/1.5041434}.

Due to the presence of dissipation (more precisely, contraction of the phase space volume), the long-time dynamics of system (\ref{1}) is determined by attractors.
Attractors can be stationary (lhs of (\ref{1}) vanishes) or non-stationary. 
Stationary attractors can be obtained from Eq.~(\ref{1}) by setting its left-hand-side to zero and solving the remaining nonlinear algebraic equations for the amplitudes $A,B,C,D$. 
In addition attractors can be symmetric (they respect the above reflection symmetry) or asymmetric. Asymmetric attractors have $D(t\rightarrow\infty)\neq 0$ and can well be stationary. The current $\bar{j}=\gamma |D|^2$ across the rhomb is nonzero for asymmetric attractors.
Symmetric attractors correspond to $D(t\rightarrow\infty)=0$ and can not be stationary.  
The current $\bar{j}=\gamma |D|^2$ across the rhomb vanishes for symmetric attractors.
The time evolution of $C(t)$, $A(t)$, and $B(t)=-A(t)$ amplitudes  of symmetric attractors turns periodic or even quasi-periodic 
\footnote{Attractors of this type were discussed earlier in Refs.~\cite{Raya15, Lled20} with respect to exciton-polariton dynamics in two coupled micro-cavities. If we associate the $A$ and $B$ sites of our system with these two cavities then, by driving the $C$-site, we pump the antibonding  mode of the $A-B$ system but it is the bonding mode which is subject to decay.}. 
For $\Omega < \Omega_{cr} \approx 0.9$ we found two symmetric attractors s1 and s2, with s1 smoothly tuned into an empty state $A=B=C=0$ for $\Omega \rightarrow 0$ and s2 keeping finite constant amplitudes $A,B,C$ in that limit. The attractor s1 turns unstable for $\Omega > \Omega_{cr}$, while attractor s2 stays stable. 
An asymmetric stationary conducting attractor a1 exists for $\Omega > \Omega_{cr}$.

\begin{figure}[t]
\includegraphics[width=8.5cm,clip]{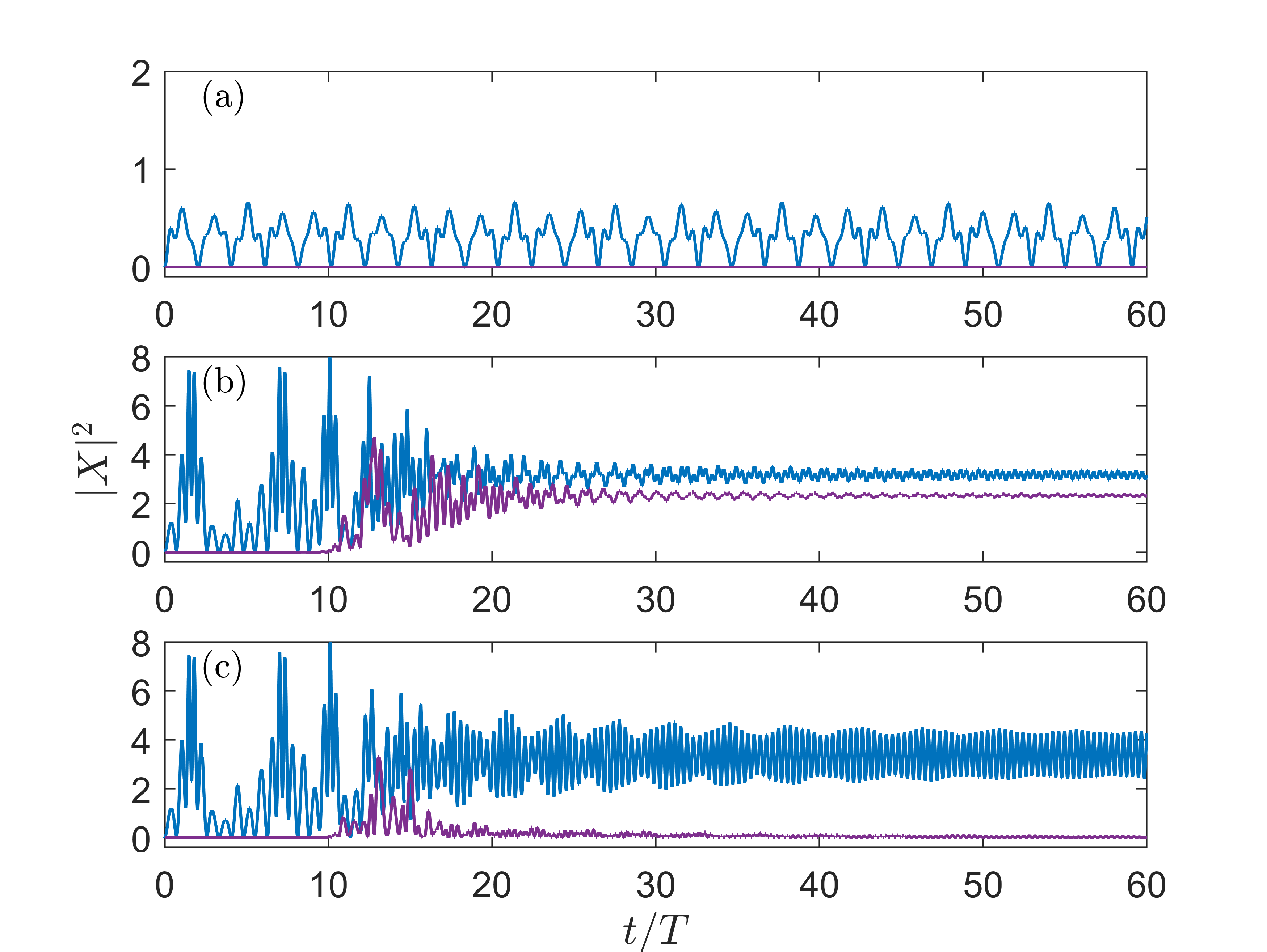}
\caption{
Relaxation of (\ref{1}) to various attractors with initial condition $|X|\equiv |A|=|B|=|C|=|D|=0.001$ and random initial phases. The plots show $|C(t)|^2$ (blue, top line) and $|D(t)|^2$ (purple, bottom line). 
(a) 
Relaxation to the symmetric nonstationary attractor s1 for $\Omega=0.85$. (b)  Relaxation to the asymmetric stationary attractor a1 for $\Omega=1.5$. 
(c) Relaxation to the symmetric nonstationary attractor s2 for $\Omega=1.5$ with a tiny change in the initial phase values as compared to case (b). 
For both (b) and (c) the amplitude $D(t)$ grows exponentially in the interval $0<t<10$, all amplitudes show chaotic dynamics during $10<t<20$, and the asymptotic regime 
is obtained for $t\gg 20$. 
Parameters are $\Delta=-0.5$, $g=0.5$, $\gamma=0.2$.}
\label{fig2}
\end{figure}

\begin{figure}
\includegraphics[width=8.5cm,clip]{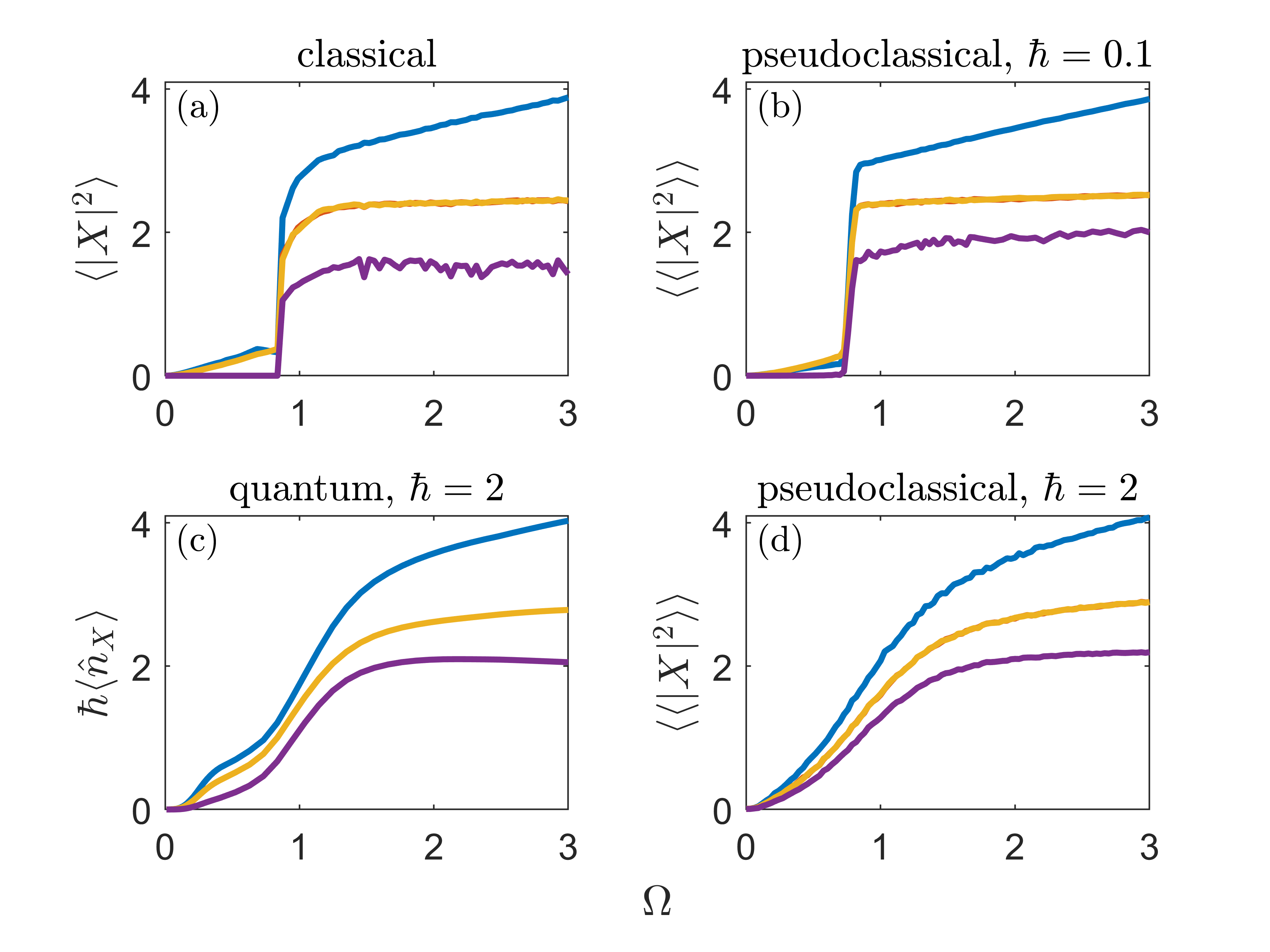}
\caption{
Left column:
(a)The mean squared amplitudes of the classical oscillators in the stationary regime versus $\Omega$, averaged over 1008 random initial phase conditions with $|X|=0.001$. 
(c) The mean populations of the rhomb sites for $\hbar=2$ in the quantum case versus $\Omega$.
Right column: 
The mean squared population of the rhomb sites for the pseudoclassical approach, (b) $\hbar=0.1$ and (d) $\hbar=2$. 
For all cases $|C|^2$ - blue curves (top), $|A|^2$ and $|B|^2$ - yellow curves (middle), $|D|^2$ - purple curves (bottom). The other  parameters are $\Delta=-0.5$, $g=0.5$,  and $\gamma=0.2$. The number of different realizations of the stochastic force in the pseudoclassical approach is 1008.
}
\label{fig3}
\end{figure}

{\em Conductance Transition.}
Let us analyze the transport across the rhomb as the function of the Rabi frequency $\Omega$.  For small $\Omega$ the steady-state response of the system to the external driving is the attractor s1 featuring small quasi-periodic oscillations with  $B(t)=-A(t)$ and $D(t)=0$ (see Fig.\ref{fig2}(a)). However, if we increase $\Omega$ above the critical value $\Omega_{cr}$ this quasi-periodic trajectory becomes unstable and any tiny perturbation (the numerical round-error suffices) leads to an exponential growth of the $D$-amplitude. The regime of exponential instability  is  followed by some transient regime of chaotic dynamics, where all amplitudes show irregular oscillations, see Fig.~\ref{fig2}(b,c). The transient chaotic regime is then relaxing into a steady-state regime a1 (Fig.\ref{fig2}(b)) or s2 (Fig.\ref{fig2}(c)). The exact attractor choice depends on the fine details of the transient state and is again affected by tiny details.  Using an ensemble of initial conditions with the absolute values of the complex  amplitudes  equal to 0.001 and random phases, we calculate the mean values of the squared amplitudes and plot them in Fig.~\ref{fig3}(a). One clearly identifies the critical driving magnitude $\Omega_{cr}\approx0.9$ and this value coincides with the transition from regular to chaotic dynamics in the Hamiltonian counterpart ($\gamma=0$) of the considered system.   Let us also mention that  from the viewpoint of a laboratory experiment the crossing of the chaos border is similar to some phase transition where the system is conducting for $\Omega>\Omega_{cr}$  and insulating for  $\Omega<\Omega_{cr}$.

{\em Quantum.}
We proceed with the quantum dynamics.  In the numerical simulations we evolve the density matrix $\widehat{{\cal R}}(t)$ for initial conditions corresponding to the empty system and calculate the single particle density matrix $\hat{\rho}(t)$
\begin{equation}
\label{b3}
\rho_{\ell,m}(t)={\rm Tr}[\hat{a}^\dagger_\ell \hat{a}_m \widehat{{\cal R}}(t)] \;.
\end{equation} 
The diagonal elements of this matrix give the population of the rhomb sites while the off-diagonal elements determine the current between the rhomb sites.
We also mention the density matrix symmetries  $\rho_{A,A}(t)=\rho_{B,B}(t)$, $\rho_{C,B}=-\rho_{C,A}$, and $\rho_{A,B}<0$, which follow from the $\pi$-flux symmetry of the quantum Hamiltonian. 
Evolving the system for long enough time,  we find the stationary population of the rhomb sites, see Fig.~\ref{fig3}(c). Note the reasonable agreement between the quantum and classical results for $\Omega>\Omega_{cr}$ and the strong discrepancy for a small $\Omega$. Unlike the classical system, the quantum system remains  conducting for $\Omega<\Omega_{cr}$. 
\begin{figure}
\includegraphics[width=7.5cm,clip]{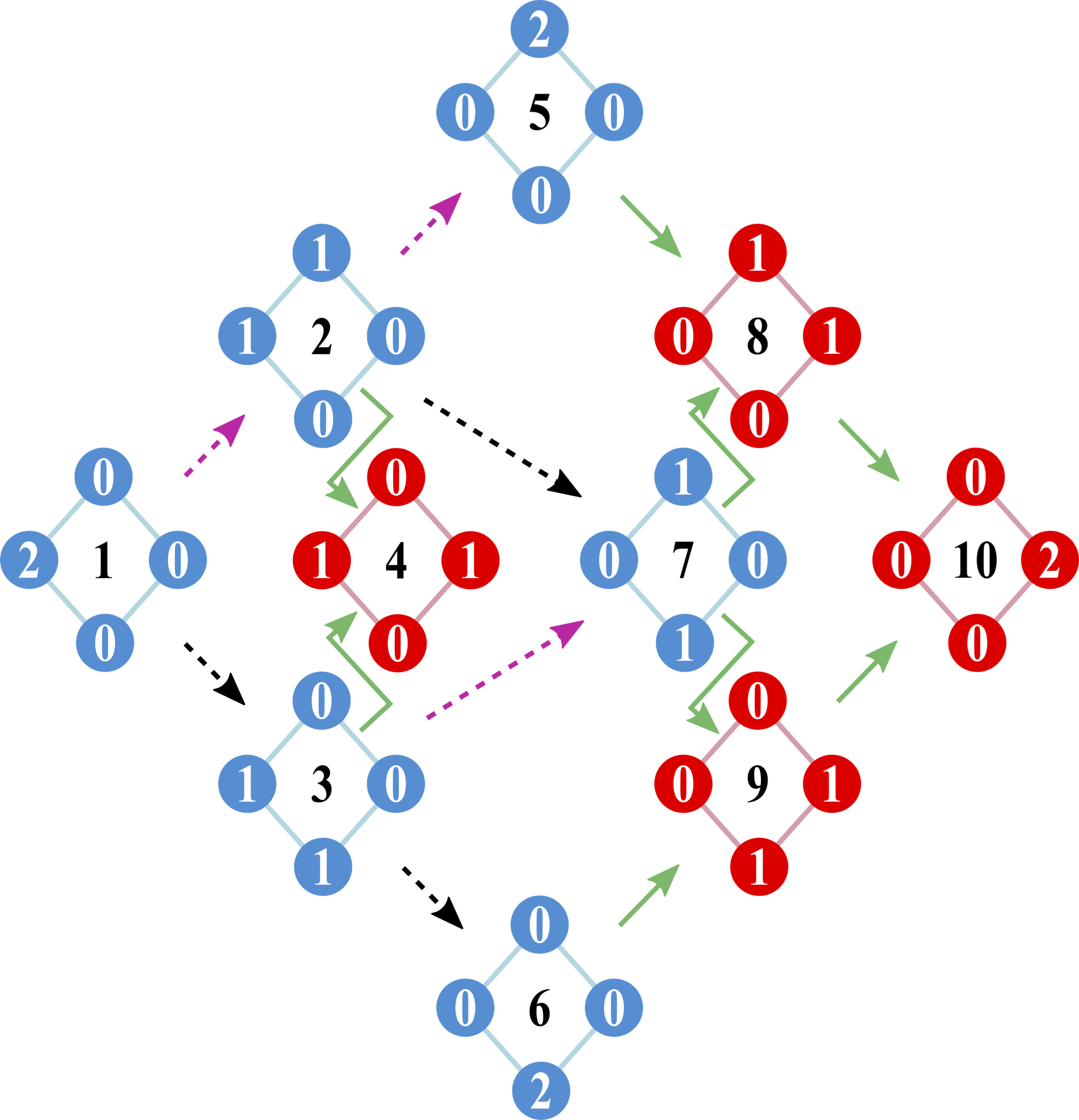}
\caption{Transition diagram for two particles $N=2$. The rhombs present the ten different Fock states which are labeled by the index  $j=1,\dots,10$ in the center of each rhomb. The numbers of particles on the individual sites of a rhomb in a given Fock state are indicated in the blue and red circles.The dashed red and black arrows correspond to matrix elements $J=\pm i$ and the green arrows to $J=1$. If $g=0$ the red-colored Fock states cannot be populated because of complete destructive interference between different paths connecting these states with the initial state $j=1$.}
\label{fig4}
\end{figure}

To explain the absence of a sharp conductance transition at a critical $\Omega_{cr}$ in the quantum case we will consider the system dynamics in Fock space. 
Let us consider for the moment $\gamma=0$ and $\Omega=0$, and an initial state of the system with a non-zero population of the $C$-site only.  
The Hamiltonian dynamics will propagate this state into other Fock states. Fig.~\ref{fig4} shows the transition diagram for two particles $N=2$ where different arrows mean different matrix elements $J$. For the noninteracting case $g=0$ the $\pi$-flux symmetry of the Hamiltonian induces destructive interference for the Fock states with non-zero occupation of the $D$-site, preventing our initial Fock state from leaking particles into the $D$-site. This holds true for larger $N$ as well \footnote{This follows from the fact that for $g=0$ the single-particle density matrix of $N$ identical particles coincides with the density matrix of a single particle.}.  Nonzero interactions $g \neq 0$ will detune and remove the destructive interference conditions. In the considered case $N=2$ these interactions change the equation of motion for the occupation amplitudes of the 5th and the 6th Fock states, which leads to a population of the symmetry protected Fock states.
 We mention, in passing, that for $N=2$ the discussed interaction-induced destruction of  Aharonov-Borm caging was confirmed experimentally in the recent  experiment \cite{Martinez2023} with super-conducting circuits.

{\em Pseudoclassical.}
The above presented results prove the quantum system to be  always conducting as soon as $g\ne0$. Thus, the insulator-to-conductor phase transition is a particular feature of the classical system. However, since the transition from the quantum to classical realms is continuous in $\hbar$, one can find a signature of this classical phase transition in the quantum dynamics if $\hbar\ll 1$. To quantitatively address this problem we resort to the pseudo-classical approximation.
The pseudo-classical approximation substitutes the master equation for the reduced density matrix by the Fokker-Planck equation for the classical distribution function $f=f({\bf a},{\bf a}^*,t)$  where $a_\ell$ and $a_\ell^*$  are the pairs of canonically conjugated variables ($\ell=1,\ldots,L$),
\begin{equation}
\label{c1}
\frac{\partial f}{\partial t}=\{H,f\} 
+\frac{\gamma}{2}\left( \frac{\partial (a_L f)}{\partial a_L} + {\rm h.c.}\right) 
+\frac{\gamma\hbar}{2} \frac{\partial^2 f}{\partial a_L \partial a_L^*} \;,
\end{equation}
where $\{\ldots,\ldots\}$ denotes the Poisson brackets and $H$ is the classical counterpart of the Bose-Hubbard Hamiltonian  (\ref{b2}). The first term in the right-hand-side of this equation corresponds to the Hamiltonian evolution of the system, the next term describes the contraction of the phase space volume, and the last term is the quantum correction to the classical Fokker-Planck equation \cite{119}.  This diffusion term is of special importance because it restricts contributions of the other quantum corrections to a value of the order of $\hbar^2$ \cite{30}.    
Unfolding the Fokker-Planck equation (\ref{c1}) into the Langevin equation, one arrives to Eq.~(\ref{1}) with an additional stochastic term $\sqrt{\hbar\gamma/2}\xi(t)$. Here $\xi(t)$ is the $\delta$-correlated complex white noise. Then the elements of the single-particle density matrix are found by averaging the solution of the Langevin equation over different realizations of $\xi(t)$, for example, $\rho_{A,B}(t)=\langle A^*(t)B(t) \rangle$. The results of the pseudoclassical approach are shown  in the right column  in Fig.~\ref{fig3}. It is seen that this approach fairly reproduces the exact quantum result for $\hbar=2$ and indicates the convergence toward the classical result for smaller $\hbar$.

{\em Conclusion.}
To summarize, we analyzed the  transport of interacting Bose particles across the rhombic lattice (diamond chain) by employing the classical (mean-field), pseudoclassical, and quantum approaches.
Within the classical approach the considered problem reduces to the dynamics of coupled nonlinear oscillators, where the first oscillator in the lattice is driven by an external field and the last oscillator is subject to friction. Then the system is insulating if the stationary amplitude of the last oscillator is strictly zero and conducting if it is finite.   Using the driving  strength $\Omega$ as the control parameter we found the system to be insulating up to  a critical $\Omega_{cr}$. When exceeding this critical value, the system becomes unstable, and the transient dynamics of the coupled oscillators turns chaotic. The final state of the system is either conducting or insulating with probability $P$ and $1-P$, respectively. 
We found that the discussed probability crucially depends on the system parameters, in particular, on the dissipation constant $\gamma$. In brief, for fixed $\Omega > \Omega_{cr}$  this probability monotonically decreases from unity to zero in the interval $0 < \gamma <0.6$. Thus there is no conducting state for $\gamma > 0.6$ (at least, in the considered interval of  $\Omega$). We also studied the many-rhomb system where, to be closer to laboratory experiments, we included small dissipation  $\tilde{\gamma}=0.01$ to all system sites. Importantly, the classification of attractors into symmetric (insulating) and asymmetric (conducting) remains valid. However, for $M>1$ one finds a zoo of possible attractor combinations. For example, for $M=3$ there are situations where the first  and second rhombs relax to asymmetric attractors but the third rhomb to the symmetric attractor. Yet, the common feature with the single-rhomb system is that the classical limit shows a sharp transition from an insulating to a conducting state at some $\Omega_{cr}$.

Within the quantum approach we numerically solved the master equation for the density matrix of interacting bosons.  We found the stationary populations of the lattice sites as a function of the Rabi frequency $\Omega$ and compared them with the classical prediction. 
While there is qualitative agreement with the classical results, one striking difference is that the quantum system conducts even below the threshold $\Omega < \Omega_{cr}$, and the sharp transition from the classical case turns into a smooth crossover for the quantum case. The reason for that is that the destructive interference is destroyed for interacting particles in the quantum case. However, we showed that the crossover sharpens back to a classical transition by using a pseudoclassical method  upon increasing the number of particles, or decreasing the effective Planck constant $\hbar$.

Acknowledgements. We thank A. Andreanov for useful discussions. This work was supported by the Institute for Basic Science,
Project Code (Project No. IBS-R024-D1). 
PSM acknowledges financial support of the Ministry of High Education and Science of the Russian Federation through grant FSRZ-2023-0006. 

\bibliography{mybib,sergejflach}

\end{document}